\newif\ifarxiv
\arxivfalse 
\ifarxiv
  \PassOptionsToClass{nonacm}{acmart}
\fi
\documentclass[sigconf, balance=false]{acmart}

\usepackage{multirow}
\usepackage{xspace}

\usepackage{listings}
\AtBeginDocument{%
  \providecommand\BibTeX{{%
    \normalfont B\kern-0.5em{\scshape i\kern-0.25em b}\kern-0.8em\TeX}}}

\ifarxiv
  \setcopyright{none}
  \renewcommand\footnotetextcopyrightpermission[1]{}
\else
  \copyrightyear{2026}
  \acmYear{2026}
  \setcopyright{cc}
  \setcctype{by}
  \acmConference[CHI '26]{Proceedings of the 2026 CHI Conference on Human Factors in Computing Systems}{April 13--17, 2026}{Barcelona, Spain}
  \acmBooktitle{Proceedings of the 2026 CHI Conference on Human Factors in Computing Systems (CHI '26), April 13--17, 2026, Barcelona, Spain}
  \acmPrice{}
  \acmDOI{10.1145/3772318.3790742}
  \acmISBN{979-8-4007-2278-3/2026/04}

  \makeatletter
  \def\@ACM@copyright@check@cc{}
  \makeatother
\fi

\definecolor{brown}{rgb}{0.59, 0.29, 0.0}
\definecolor{darkgray}{rgb}{0.59, 0.59, 0.59}
\definecolor{tablegray}{gray}{.9}
\definecolor{historypalettepurple}{HTML}{5656b0}

\newcommand\highlight[1]{\textcolor{red}}

\usepackage[utf8]{inputenc}
\usepackage{diagbox}
\usepackage{colortbl}

\usepackage{tabularx}

\usepackage{color}
\usepackage{enumitem}
\usepackage{mathtools}
\usepackage{commath}

\usepackage{amssymb}
\usepackage{pifont}

\newcommand{\customtilde}{{\raise.17ex\hbox{$\scriptstyle\sim$}}}

\newcommand{\eg}{\textit{e.g.},\xspace}
\newcommand{\ie}{\textit{i.e.},\xspace}

\usepackage{xparse}

\newcommand{\system}{HistoryPalette}

\definecolor{changeNoteColor}{RGB}{158, 158, 219}
\newcommand{\changenote}[1]{#1} 

\begin{document}

\title{\system: Supporting Exploration and Reuse of Past Alternatives in Image Generation and Editing }

\author{Karim Benharrak}
 \orcid{0009-0002-3279-5664}
 \email{karimbenharrak@berkeley.edu}
\affiliation{%
   \institution{University of California, Berkeley}
   \city{Berkeley}
   \state{California}
   \country{USA}
 }

\author{Amy Pavel}
 \orcid{0000-0002-3908-4366}
 \email{amypavel@eecs.berkeley.edu}
 \affiliation{%
   \institution{University of California, Berkeley}
   \city{Berkeley}
   \state{California}
   \country{USA}
 }


\begin{abstract}
Creative tasks require creators to iteratively produce, select, and discard potentially useful ideas.
Now, creativity tools include generative AI features (e.g., Photoshop Generative Fill) that increase the number of alternatives creators consider through rapid experiments with prompts and random generations. 
Creators use tedious manual systems for organizing their prior ideas by saving file versions or hiding layers, but they lack the support they want for \rev{reusing prior alternatives in personal work or in communication with others}.
We present \system, a system that supports exploration and reuse of prior designs in generative image creation and editing.
Using \system, \rev{creators and their collaborators} explore a ``palette'' of prior design alternatives organized by spatial position, topic category, and creation time.
\system{} enables creators to quickly preview and reuse their prior work.
In \rev{creative professional and client collaborator user studies}, participants generated and edited images by exploring and reusing past design alternatives with \system{}. 
\end{abstract}





\newcommand{\rev}[1]{#1} 

\maketitle

\begin{figure*}
  \includegraphics[width=\textwidth]{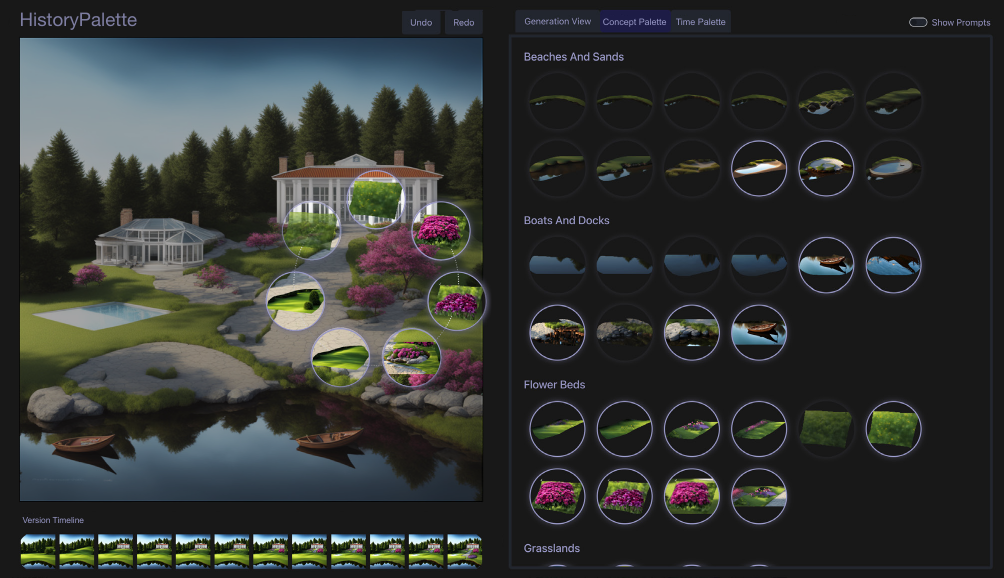}
  \caption{\system{} enables rapid exploration and reuse of past design alternatives (circles). Users use the generation view to edit their image by selectively generating new content onto the canvas. \system{} captures design alternatives created through edits and stores them for later exploration. Using the palettes (Position Palette, Concept Palette, and Time Palette) users can explore past design alternatives by rapidly previewing them in the place they originally occurred. \system{} groups alternatives in clusters and classifies previous unsuccessful generations to be bad past alternatives (grayed-out alternatives)}
  \Description{An interface of a digital tool called 'HistoryPalette,' displaying a landscaped scene with a central mansion surrounded by trees, a garden, and a pond with small boats. On the left, circular overlays highlight specific features in the scene, such as flower beds, grasslands, and docks. On the right, there is a panel with categorized circular thumbnails under headers like 'Beaches and Sands,' 'Boats and Docks,' 'Flower Beds,' and 'Grasslands,' showing variations or options for these elements. The bottom features a timeline ('Version Timeline') of different versions of the landscape scene. The interface includes 'Undo' and 'Redo' buttons on the left side and toggles on the right side for 'Generation View', 'Concept Palette' and 'Time Palette.'}
  \label{fig:interface}
\end{figure*}

\section{Introduction}
\changenote{
Creators, such as digital artists, create original pieces from existing source material when compositing images or creating collages.
While visual elements to use in compositions were traditionally gathered from external sources like the web or asset libraries, text-to-image generation tools such as Midjourney~\cite{midjourney}, Dall-E~\cite{dalle}, and Photoshop Generative Fill~\cite{adobephotoshop} enable rapid exploration of visual elements directly in-place thus reduce time, effort and knowledge required to create high-fidelity image compositions.
}
However, prompt-based generation trades direct control for underspecified text prompts such that creators use frequent trial-and-error to reach their final outcome. 
Recent work helps creators improve their prompts by suggesting styles~\cite{liu2022opal,wang2024promptcharm}, providing refinements~\cite{brade2023promptify, feng2023promptmagician}, and adjusting model attention to prompt text~\cite{wang2024promptcharm}, but each generation still takes time, costs computing power, and yields uncertain outcomes. 
Such costs compound in projects that require many generated elements \changenote{(\textit{e.g.}, scene and image compositions, collages)} as creators try multiple prompts per element and each prompt generates multiple alternatives.
While discarded generations do not suit a creator's immediate needs, they represent potentially valuable alternatives for future use. 
Rather than improving prompts, our work seeks to improve generative workflows \changenote{in existing commercial tools (\eg Photoshop~\cite{adobephotoshop}, GIMP~\cite{GIMP}} by supporting the \textit{exploration} and \textit{reuse} of prior generations.

Managing creative histories is a longstanding challenge across all creative work as creators try, discard, and reuse ideas to reach their final composition~\cite{schon2017reflective,shneiderman2007creativity}. 
These histories provide \textit{an ordered record} of the creator's process and a \textit{collection of alternatives} for communication or reuse~\cite{shneiderman2007creativity,sterman2022CreativeVersionControl}
For example, a digital artist may use their history to show a collaborator how they created a portrait, or revisit their previous portrait backgrounds to show alternatives~\cite{sterman2022CreativeVersionControl}.
Currently creators must manually manage alternatives by using undo/redo logs, saving separate files~\cite{sharmin2009understanding,li2021we,o2018charrette}, or using layers~\cite{adobephotoshop,figma}.
Such manual approaches are error-prone as creators forget to save, struggle to locate, or inadvertently delete alternatives --- 
thus creators can not access the alternatives they want to find~\cite{sharmin2009understanding,sterman2022CreativeVersionControl}.
Prior research makes histories easier to navigate for personal and collaborative use
~\cite{klemmer2002web,grossman2010chronicle,Angert2023SpellburstAN,dang2023worldsmith,Rawn2023} by providing timeline annotations~\cite{klemmer2002web,grossman2010chronicle} and preserving branching~\cite{dang2023worldsmith,Angert2023SpellburstAN,Rawn2023} of project histories.
Both current practice and prior research primarily order creative histories chronologically.
While time-based organization supports review of the creative process, finding specific alternatives (\textit{e.g.}, previously created trees or backgrounds) requires manual review of the entire project history that can contain hundreds to thousands of edits.

New prompt-based workflows intensify and also provide opportunities to solve history management challenges.
While the trial-and-error of prompt-based workflows creates more alternatives to manage, prompt-based workflows also provide semantic information absent in traditional editing. Each prompt-based edit provides a semantically meaningful history segment (\textit{e.g.}, a tree vs. a set of strokes) along with two semantically meaningful handles (a target region and a prompt). Such handles may better match how creators and their collaborators think about the work --- in terms of what content appears where, rather than when it was created.
Recent advances in generative models also offer opportunities for history management and reuse as modern vision language models can filter out low quality edits by comparing the prompt text to the generated result, and generation-based smoothing can incorporate past alternatives into current work.
We investigate the potential to use such advances for history organization to support the exploration and reuse of the project history in generative workflows.

Towards this goal, we contribute \textit{\system{}}, a system for exploring and reusing prior alternatives \changenote{to support image composition via generative in-painting} (Figure~\ref{fig:interface}).
\system{}'s Generation View supports image generation workflows similar to Photoshop Generative Fill~\cite{adobephotoshop} where creators iteratively select a region of the image to change, then write a prompt to generate a new image that fits seamlessly within that region.
At each editing step, \system{} records a version of the entire image (i.e. \textit{version}) and the latest text-to-image generation edit consisting of a region, image, and prompt (i.e. a partial-project \textit{alternative}).
\system{} then provides four history views: a \textit{position palette} that lets creators access alternatives at the current position, a \textit{concept palette} lets creators access alternatives clustered by prompt themes, a \textit{time palette} that lets creators access alternatives by time, and a linear version timeline that lets creators access the version history of the entire project.
While the position palette and concept palette are unique to \system{}, the time palette and version timeline more closely mimic temporal organization in existing history management.
Creators can hover over any alternative or version to preview it in place, or click to paste it \changenote{anywhere} into their project, optionally rasterizing the image to smooth artifacts.
\system{} further supports history filtering via automatic removal of image generation failures using a vision language model, as well as manual removals/favorites for later use or sharing with clients.

We conducted a technical evaluation,
\textit{a creative professional study} with three participants (41 years combined experience) to explore the potential of \system{} in creative workflows of creative professionals, and a \textit{client collaborator study} with eight participants to explore the potential for efficient history access as a collaboration tool.
Our technical evaluation demonstrated that while 18.8\% of prompts generated errors, our filter identified them with 93\% accuracy (86\% precision, 76\% recall). 
Our user studies demonstrated how semantic organization of generative histories can benefit both creators and their collaborators.
Across both studies, all participants used our semantic organization methods (position and concept palette) more often than the traditional time-based organization (time palette).
Professional participants used prior alternatives to inspire new ideas and test alternatives, and expressed enthusiasm about using \system{} in the future for open-ended and collaborative work.
Our client collaborator evaluation revealed that participants used \system{}'s prior generations ($\mu=11.88$ times, $\sigma=6.09$ times) more often than they used new generations ($\mu=4.62$ times, $\sigma=1.73$ times) to edit the image, reporting that they preferred the efficiency and control of reusing prior generations. 
Client collaborator participants also expressed excitement to use \system{} for providing feedback, learning expert prompt strategies, and exploring personal preferences.
Our findings demonstrate the potential for semantic organization to encourage reuse in generative workflows.
\changenote{We conclude with implications for prompt-based workflows and collaborative use and opportunities to extend the utility of our palettes to encompass more interaction types and history structures.}
\section{Background}
\system{} supports creators in image generation workflows using their prior history as a source of inspiration and reuse.
Our work relates to prior work in history reuse for authoring and communication, using external reference examples for ideation and reuse, and interfaces for improving text-to-image generation. 

\subsection{History Keeping and Reuse for Authoring}
Creative work requires creators to ideate, implement, discard, reuse, and recombine ideas until they reach their final composition in a ``reflective conversation'' with their work~\cite{schon2017reflective}. As creators often discard many useful ideas in their process, a longstanding goal for creativity support tools has been to enable ``rich history keeping''~\cite{shneiderman2007creativity} to support creators in recording what they have tried, comparing alternatives, revisiting and reusing prior versions, and communicating variations to collaborators~\cite{shneiderman2007creativity}.
Currently, practitioners manually store rich histories of their projects for later inspiration and reuse via files, copies or layers~\cite{sterman2022CreativeVersionControl,sharmin2009understanding}. 
Layers can maintain partial-project alternatives~\cite{terry2002recognizing} but can be difficult to organize or explore on large projects.
For example, with Photoshop Generative Fill, each generation produces a new layer with three alternatives hidden inside the layer (defaults to a chronological order) such that finding and reusing prior generations remains challenging. In addition, creators often need to flatten their layers to improve subsequent generations thus discarding their alternatives. We aim to support on-demand exploration and reuse of the history of alternatives.

Prior work has explored using history for exploration and reuse in domains where code or parameters generate the image (\textit{e.g.}, for 2D creative coding, generative art, and image manipulation)~\cite{Subbaraman2023ForkingAS,Angert2023SpellburstAN,Rawn2023,Zaman2015,terry2004variation}. 
For example, Parallel Paths~\cite{terry2004variation} added variations in place during parameter adjustment for image manipulations (\textit{e.g.}, degrees of application of the whirl filter).
More recently Quickpose~\cite{Rawn2023}, Spellburst~\cite{Angert2023SpellburstAN}, GEM-NI~\cite{Zaman2015}, and Subbaraman et al.~\cite{Subbaraman2023ForkingAS} explore approaches for branching and remixing alternatives in creative code and generative design.
\changenote{While prior work has shown that parallel design exploration is beneficial~\cite{terry2004variation, Rawn2023, Angert2023SpellburstAN, Zaman2015, Subbaraman2023ForkingAS}, existing systems only support iterative design exploration workflows.}
Furthermore, prior work focuses on parameter- or code-based authoring and does not yet address creative tasks that operate primarily on pixels rather than parameters (\textit{e.g.}, digital painting, photo manipulation, video editing). 
Pixel-based editing lacks a clear structure for identifying salient temporal and spatial regions for replacement or reuse (\textit{e.g.}, what set of edits or brush strokes make up a salient scene lighting change). 
We explore the potential for history reuse in pixel-based \changenote{image} editing by investigating prompt-based workflows that offer new structure (\textit{e.g.}, prompts and regions).
We use the metaphor of a \textit{history palette} inspired by Sterman et al.'s formative work identifying that creators desire to use their history as a ``palette of materials'' for reuse~\cite{sterman2022CreativeVersionControl}.

\subsection{History Reuse for Communication}
For a broad range of design domains, researchers have explored how to parse project histories to use as a learning and communication resource~\cite{klemmer2002web,grabler2009generating,grossman2010chronicle,pavel2013browsing,kong2012delta}.
Klemmer et al.'s~\cite{klemmer2002web} wall display interface records design process history 
then presents it in a timeline (a time-ordered set of version thumbnails) that can be filtered (\textit{e.g.}, by author)
to support design collaboration. 
Chronicle~\cite{grossman2010chronicle}, Grabler et al.~\cite{grabler2009generating}, Delta~\cite{kong2012delta}, and Sifter~\cite{pavel2013browsing} all support novices learning from expert histories by allowing novices to browse annotated linear timelines of edit histories~\cite{grossman2010chronicle}, converting edit histories into text tutorials~\cite{grabler2009generating}, or comparing multiple expert methods~\cite{kong2012delta,pavel2013browsing}.
All such prior work seeks to address learning the process and thus supports chronological navigation, but we aim to use the history to support communication via the review of potential alternatives or direct history-based adjustments by a design collaborator. 

\subsection{Reference Examples for Ideation and Reuse}
In the early stages of the creative process, practitioners look for reference examples, understand the existing space of prior designs, and then modify the designs to generate new ideas~\cite{herring2009getting,herring2009idea, sharmin2009understanding}.
Searching for relevant examples can be challenging as traditional text-based search supports searching by style or layout.
Thus, prior research has explored domain-specific reference example search tools for web design~\cite{lee2010designing,hartmann2010d}, mobile interfaces~\cite{bunian2021vins,huang2019swire}, 3D models~\cite{funkhouser2003search,Matejka2018DreamLE}, fashion~\cite{laenen2018web,park2019study}, visual metaphors~\cite{Kang2021MetaMapSV}, and more. 
Once practitioners find reference examples, recombining the examples can inspire new designs~\cite{Chilton2019VisiBlendsAF, Kerne2000CollageMachineAI,choi2023creativeconnect, kumar2011bricolage,warner2023interactive,Kim2022MixplorerSD}. 
Prior work lets users combine reference examples via style transfer~\cite{kumar2011bricolage,warner2023interactive}, blending images~\cite{Chilton2019VisiBlendsAF, wang2023popblends}, collaging web elements~\cite{Kerne2000CollageMachineAI}, and merging garden designs~\cite{Kim2022MixplorerSD}.
Prior work also explored generating rather than referencing existing examples to support designers in improving their designs~\cite{Swearngin2020,o2015designscape,Ivanov2022MoodCubesIS}. For example,
Scout~\cite{Swearngin2020} and DesignScape~\cite{o2015designscape} generate design alternatives using visual design principles and user-created constraints.
Others' examples and design refinements are valuable, but as practitioners work they also build a rich library of reference examples in their own project history. We explore the use of practitioners' own history for inspiration and reuse.

\subsection{AI Image Generation and Editing}
Text-to-image generation (\textit{e.g.}, with GANs~\cite{Qiao2019LearnIA, Qiao2019MirrorGANLT, Xia2021TediGANTD}, diffusion models~\cite{ruiz2023dreambooth}) has rapidly improved over the past few years and is now widely available in commercial tools (\textit{e.g.}, Midjourney, DALL-E, Photoshop's Generative Fill). 
Such tools lack controllability as users must iteratively formulate text prompts and this randomness in outputs may enhance inspiration~\cite{Vimpari2023AnAT} or decrease originality~\cite{wadinambiarachchi2024effects}.
To support creators reaching their creative goals, prior work has helped users create prompts~\cite{Liu2022OpalMI,brade2023promptify,wang2024promptcharm,feng2023promptmagician}, explore the input-output space in text-to-image generation~\cite{almeda2024prompting}, and generate complex and specific layouts~\cite{dang2023worldsmith,huang2024plantography}. In this space, WorldSmith~\cite{dang2023worldsmith} which uses image generation tiles for world building is closest to our work. WorldSmith provides a history view (via a tree) that lets users branch the project to create parallel versions.
However, similar to other branching node-based histories~\cite{Subbaraman2023ForkingAS,Rawn2023, zuend2017storyversioncontrol}, an ordered tree becomes difficult to skim and browse for quick access to alternatives, especially as the project grows.
We explore how to create right-at-hand access to alternatives via a ``palette'' for history reuse.
\section{HistoryPalette}
\changenote{We designed a system to support history exploration and reuse for image composition via generative in-painting.
}


\subsection{Design Rationale}
\changenote{We designed \system{} based on three assumptions: (1) unlike pixel-level direct manipulation, prompt-based generations represent \textbf{semantically complete edits} (\ie~prompt and target region), (2) generations are reusable ~\textbf{independent alternatives} (\ie reusable without dependency conflicts), and (3) discarded generations are potential alternatives that \textbf{users want to explore and reuse later}~\cite{terry2004variation, shneiderman2007creativity, sterman2022CreativeVersionControl}.
Based on these assumptions, we designed \system{} with four main goals:
}
\\

\noindent \textbf{G1: Palettes of Alternatives for Discovery. } Creators use their past work as a ``palette of materials'' for future use~\cite{sterman2022CreativeVersionControl}. We thus create palettes that support the use of workflow primitives (\textit{i.e.}, a palette \textit{alternative} -- or, an image region with a prompt and a generation) for ease of alternative discovery and reuse. We considered three dimensions to support quickly discovering and dipping from the palettes: time, canvas position, and creative concept. 
Time provides a default index into the creator's process. Time groups together related actions~\cite{grossman2010chronicle} and supports recall for the project creator (\textit{e.g.}, ``I know I created flowers after changing the sunset'').
Time most closely mirrors the behavior that layers provide upon generation today~\cite{adobephotoshop}.
Canvas position supports creators exploring their palettes directly by clicking on the intended object in the image to see local alternatives.
As position lets creators map their intent (\textit{e.g.}, change the flowers) onto the canvas (an x,y coordinate on the flowers), it is useful when creators know what they want to change. 
Position palettes limit discovery as they must be hidden until clicked to avoid canvas occlusion, so we aim to create a palette of discoverable alternatives that preserves the relevance of position palettes.
Thus, we create concept groups with text labels to support browsing for alternatives of interest. 
Concept groups serve as a palette for creators attempting to problem solve a portion of the design (\textit{e.g.}, ``landscaping'' palette with grass, bushes, and gardens, or a ``buildings'' palette with castles and houses).
\\

\noindent \textbf{G2: Support Rapid Reuse of Alternatives.}
To let users freely explore and reuse the palette of alternatives, we aimed to support fast in-place previews of alternatives and on-demand flexible reuse. To support rapid reuse, we provide previews on hover to create a low-risk environment for experimenting with alternatives. Such experiments may support decision-making before expensive edit operations.
We then explore two options to let users place \changenote{reused alternatives}: simply pasting the alternative on the canvas, and a generative approach that smooths the image including the reused alternative to remove editing artifacts. \\ 


\noindent \textbf{G3: Provide History Filters and Favorites.} Histories de-risk exploration~\cite{shneiderman2007creativity}, so we preserve the entire history. However, the history contains a spectrum of alternatives from equally viable options to complete failures (\textit{e.g.}, a house with a different roof vs. the prompt ``house'' fails to generate a house). 
Generation quality, personal taste, and current image state all dictate the usefulness of alternatives, but ``bad'' alternatives can clutter alternative palettes. 
We aim to filter out alternatives that are not likely to be useful, then let designers further curate their history manually for personal and collaborative use~\cite{o2018charrette}.
All filtering and removal are non-destructive such that users can recover filtered alternatives in the interface. \\

\noindent \textbf{G4: Support Multiple Audiences. } Histories are useful for the creators, but also for collaborators and novices~\cite{shneiderman2007creativity,grossman2010chronicle,grabler2009generating}. While prior work supported novices to learn from the designer's history to perform the task on their own~\cite{grabler2009generating,grossman2010chronicle}, prior work has not yet explored supporting collaborators to reuse alternatives in a designer's history. If successful, collaborators can use the same abstractions provided to creators to understand the design process and considered alternatives such that they may build common ground with the designer to give useful feedback~\cite{shneiderman2007creativity}. 
Thus, we aim to support audiences beyond the creator via externally meaningful history organization and surfacing expert prompt use.

\subsection{\system{} Overview}
\changenote{
\system{} lets users compose images by iteratively generating, exploring, and reusing design alternatives.
\system{} features three main components: a \textit{generation view} for creating and editing images via inpainting; three design history palettes (\textit{concept palette}, \textit{position palette}, and \textit{time palette}) for browsing and reusing prior alternatives, and a \textit{version timeline} for browsing and reusing project-level history versions.
\system{} is designed to support non-linear design exploration workflows where users iteratively change prompts, switch between versions, and sometimes even start from new.
Unlike traditional chronological history trees that assume a single root and isolate alternatives within specific branches~\cite{Angert2023SpellburstAN, dang2023worldsmith, Rawn2023}, \system{} flattens these trees by aggregating all explored alternatives into a global repository of reusable materials.
~\system{} organizes all alternatives by semantic concepts (\textit{concept palette}) and spatial location (\textit{position palette}).
We mimic closed-source workflows such as when using Adobe Photoshop's Generative Fill~\cite{adobephotoshop}, and the \textit{version timeline} mimics chronological project history~\cite{Zaman2015}. We follow prior work in visualizing version thumbnails to support skimming~\cite{klemmer2002web}.
We implemented the time palette for simple chronological ordering as a baseline to mimic the behavior of sequentially created layers~\cite{adobephotoshop,figma}.
Our system explores how history palettes can be used in existing creativity support tools and history-keeping approaches to enable design exploration and reuse.
}



\subsection{Interface}
The \textbf{composition view} (Figure~\ref{fig:interface}, left) starts blank or with an uploaded background image and displays the current composition. Authors can add to their composition via the generation view, the palettes, or the version timeline. 

\system{}'s \textbf{generation view} includes a prompt box along with optional sliders to change generation parameters for strength, guidance, and inference steps. To use the generation view, creators draw a \textit{region} in the composition where they want to place their generation, write a \textit{text prompt}, optionally change parameters, and click submit. The generation view provides four generation \textbf{alternatives} that creators can hover over to preview in-place. Creators can click a generation alternative to add it to their composition or click undo to try again. 
All discarded and selected alternatives from the generation view are added to \system{}'s palettes of alternatives for later use. 
We run a \textbf{filter} over each generated image alternative and its prompt to filter out alternatives that are likely not useful. For example, if the creator drew a circle in the lake and prompted ``boat'' but the generation failed to add a boat, we flag the alternative as a failed generation. Failed generations are hidden by default, but users can press `shift' to display them in the palettes with a lower opacity (Figure~\ref{fig:interface}, right).
Each edit also adds a \textbf{version} of the entire composition to the version timeline. 

\system{}'s \textbf{palettes} display all generation alternatives, which creators can explore, preview, and reuse.
Users can \textit{hover} over an alternative in a palette to preview the alternative in place or \textit{click} it to add to the composition. 
Each added alternative can be transformed by dragging the alternative around the image or changing its scale.
Creators can \textbf{paste} the alternative directly in place, potentially leaving boundary artifacts due to background mismatches between the alternative and the current image composition, or \textbf{rasterize} which decreases boundary artifacts by smoothening out edges, but takes more time and occasionally changes other small composition details.

\begin{figure}[t]
    \centering
    \includegraphics[width=3.33in]{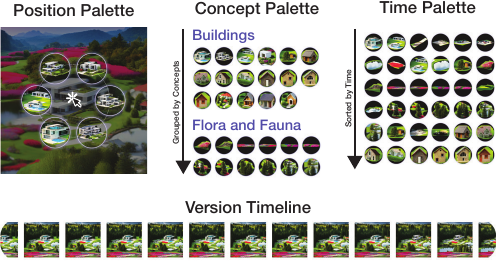}
    \caption{Overview of palettes in our system: the position palette (left), the concept palette (middle), and the time palette (right). The version timeline of full project versions appears at the bottom.}
    \label{fig:palettes}
    \Description{An image displaying an overview of three palettes and a version timeline. On the left is the 'Position Palette,' showcasing a landscape scene with circular overlays highlighting specific features, such as buildings or vegetation. In the center is the 'Concept Palette,' divided into two sections: 'Buildings' and 'Flora and Fauna,' with circular thumbnails grouped by concepts. On the right is the 'Time Palette,' showing a series of circular thumbnails arranged in rows, sorted by time progression. At the bottom, the 'Version Timeline' displays a sequence of thumbnails depicting iterations of the landscape scene.}
\end{figure}

Creators access the \textbf{position palette} by clicking on a region in the composition canvas to see other alternatives generated for that region. For example, a creator may click on a house in an image to see house alternatives generated at that position (Figure~\ref{fig:palettes}), and hover over the alternatives to preview them in place. 

Creators access the \textbf{concept palette} and \textbf{time palette} via tabs (Figure~\ref{fig:interface}, top). The concept palette groups alternatives by prompt themes (Figure~\ref{fig:interface}, left).
For example, a creator may skim the building alternatives in the concept palette (Figure~\ref{fig:palettes}) and decide to add other dwellings from the palette into their scene. The time palette orders all alternatives by recency with the most recent at the top (Figure~\ref{fig:palettes}). 

Creators can tag any alternatives as failed alternatives, and \textbf{favorite} alternatives for later (\eg to show a client during project delivery).
Creators can ``show prompts'' (Figure~\ref{fig:interface}, top right) to see what prompts created each alternative.

Finally, \system{} includes a version timeline that displays every prior version of the composition.
Creators \textit{hover} over the version timeline to preview the prior composition in the composition window, and \textit{click} the version to restore the prior version. When creators restore the prior version, the rest of the history remains. The new version is simply added at the end of the version timeline and the palettes of all prior alternatives remain available.

\subsection{Technical Methods}
Our technical pipeline consists of an image inpainting pipeline to support generative fill and rasterization, a filtering approach to hide unsuccessful alternatives, and approaches for organizing alternatives by position and concept for our palettes.

\begin{figure*}[t]
    \centering
    \includegraphics[width=\textwidth]{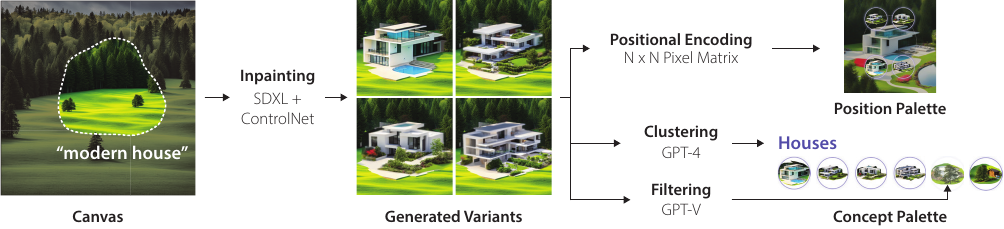}
    \caption{Users interact with the canvas by selecting a region and entering a prompt, such as ``modern house.'' \system{} presents 4 generated design alternatives per request using Stable Diffusion XL~\cite{podell2023sdxl} and ControlNet~\cite{zhang2023adding}. Each design alternative is post-processed for \system's palettes. The positional encoding organizes design alternatives by location for the position palette, while clustering by concept groups design alternatives into relevant categories like ``Houses'' in the concept palette.}
    \label{fig:pipeline}
    \Description{On the left, a 'Canvas' displays a grassy field with a dashed outline labeled 'modern house.' An arrow points to 'Inpainting,' labeled 'SDXL + ControlNet,' leading to 'Generated Variants,' which showcases four modern house designs with different architectural styles. From the variants, three arrows lead to separate processes: 'Positional Encoding (N x N Pixel Matrix)' connects to a 'Position Palette' showing a layout with circular overlays of house designs; 'Clustering (GPT-4)' connects to a labeled category 'Houses' with circular thumbnails; and 'Filtering (GPT-V)' connects to a 'Concept Palette' displaying the same house thumbnails.}
\end{figure*}

\subsubsection{Image Inpainting Pipeline}
We implemented an image inpainting pipeline to mimic existing generative fill features (\textit{e.g.}, Photoshop Generative Fill~\cite{adobephotoshop}).
Once a user selects a region on the canvas (\textit{e.g.,} draws a circle in the middle of the image) and provides a prompt (\textit{e.g.,} ``house''), we pass the canvas image, selected region, and prompt to an Stable Diffusion in-painting module~\cite{rombach2021highresolution, zhang2023adding} to edit the selected region based on the prompt.
We tested different parameters (\textit{e.g., strength, guidance, step size}) and initialize them with default values that we empirically found to work the best (see \ref{sec:appendix-parameters}). These parameters may optionally be adjusted in the Generation View.
We generate 4 generation alternatives for each prompt and selected region, similar to existing image generation features~\cite{adobephotoshop,midjourney}. When a generation is chosen, we add all 4 alternatives to the history.

When reusing prior alternatives, we let users resize, transform, or directly paste the alternative into the canvas. As the background of a reused alternative may not match the current background, we used the image-to-image refiner model from the Stable Diffusion XL pipeline~\cite{sdxlrefiner} to blend their alternatives into the new location. We use \textit{``high resolution, high quality''} as prompt and only run 7 inference steps to avoid the image changing too much.

\subsubsection{Filtering Alternatives} 
While image generations are typically successful (\ie follows the prompt), they may occasionally be unsuccessful (\eg ignores the prompt). For example, if the user selects a grass region and prompts the model to generate ``yellow flowers'' the model may ignore the prompt and generate a new patch of grass or a bush. To ensure such failures do not clutter the history palette, we classify alternatives as successful or unsuccessful. Specifically, we prompt a  multimodal LLM with vision capabilities (GPT-4V~\cite{openai2023api}) to serve as a binary classifier of successful (True) or unsuccessful (False) image generations (Appendix~\ref{sec:appendix-prompt-filtering}). We provide the model with a classification prompt, the inpainting generation (\textit{e.g.}, the mask with a generated bush and surrounding grass), and the user's inpainting prompt (\textit{e.g.}, yellow flowers). 
We run the classifier on each alternative that is newly added to the history and provide few-shot examples of successful and unsuccessful generations.

\subsubsection{Organizing and Reusing Alternatives}
To power our concept palette, we prompted GPT-4-turbo~\cite{openai2023api} (Appendix~\ref{sec:appendix-prompt-clustering}) to cluster a list of tuples of alternative indices and their prompts from the entire history (\textit{e.g.}, [0] yellow flowers, [1] bushes, [2] pink flowers) and output a JSON dictionary with the cluster names and the alternative indices that belong to the respective cluster (\textit{e.g.}, \{"flowers": [0, 2], ... \}).
We re-cluster all alternatives whenever a new alternative is added to the history.
For our position palette, we store a $N\times N$ matrix representing a $N\times N$ pixel image where each entry contains a chronologically ordered list of alternatives for the respective pixel based on the mask of the alternative. For each new alternative, we update this matrix. When a user clicks on the image to open the position palette, we display all alternatives of the respective matrix entry. If the alternative occupies the entire image, we do not add the alternative to the matrix but instead display the alternative in the top left of the image to indicate that it is a global rather than a local alternative.
For the time palette, we store a timestamp of the alternative's creation for chronological ordering.


\subsection{Implementation}
We implemented \system{}'s frontend using React and Fabric.js. We used Python Flask API for our backend. The image inpainting and rasterization models were hosted on Huggingface inference endpoints.
The backend logs six action events: image inpainting, rasterization, copy-pasting, reverting to an old project state, and undo and redo actions.
All actions are saved in a CSV file for each project.
We evaluate and cluster each action using GPT-4~\cite{openai2023api}.
\subsection{Technical Evaluation}

We analyzed the three design projects created in the creative professionals study to assess the quality of our system's inpainting generations and how well we filter out unsuccessful generations.

We extracted all image generation prompts with their corresponding sets of four generated alternative images across all projects.
Each project included 11 ($\sigma=2.94$) ~prompts for a total of 132 prompt and image pairs across all projects.
We then reviewed and annotated each pair as ``unsuccessful'' if it: (1) did not generate an object, (2) generated the wrong object (\textit{e.g.}, ``yellow flowers'' led to a bush), (3) generated the object but placed it illogically (\textit{e.g.}, a ``horse'' prompt generated a horse head in the grass), or (4) the generation created artifacts along with the object. 
Our annotation revealed that 18\% of the pairs were unsuccessful. 
We ran our filtering approach on the prompt-image pairs and evaluated against our ground truth annotations. 
Our filtering approach achieved an accuracy of 0.93, a precision of 0.86, and a recall of 0.76. We aimed for lower recall than precision to conservatively hide alternatives.
To cope with remaining errors we let users manually hide/unhide alternatives or toggle to display hidden alternatives.

\changenote{
We also measured delays in our image inpainting pipeline that relied on Stable Diffusion XL~\cite{rombach2021highresolution}. Within the sessions of creative professionals, new generations took an average of 29.51 seconds ($\sigma=0.71$, max = 31), while rasterizing a reused alternative took 12.89 seconds ($\sigma=8.26$, max = 48). We compare the delays in \system{} to Photoshop Generative Fill~\cite{adobephotoshop}. We measured the generation delay for 10 individual inpainting actions when using Adobe's Firefly 3 model ($\mu$ = 13.7s, $\sigma$ = 0.21, max = 17.2) and Nano Banana Pro ($\mu$ = 25s, $\sigma$ = 4.62, max = 64).
Thus, while \system{}'s delays are slightly higher, they remain comparable to delays in existing commercial tools.
}

\begin{table*}[t]
\resizebox{\textwidth}{!}{%
\begin{tabular}{llllllllllllllll} 
\toprule
\textbf{\textbf{ID}} & \textbf{\textbf{Time}} &  & \multicolumn{5}{l}{\textbf{Preview}} &  & \multicolumn{5}{l}{\textbf{Reuse}} &  & \textbf{\textbf{\textbf{\textbf{Generations}}}} \\ 
\midrule
 &  &  & \multicolumn{4}{l}{Palettes} & Version~Timeline &  & \multicolumn{4}{l}{Palettes} & Version~Timeline &  &  \\
 &  &  & Position & Concept & Time & Total &  &  & Position & Concept & Time & Total &  &  &  \\ 
\midrule
E1 & 69 mins &  & 80 & \textbf{155} & 0 & 235 & 72 &  & 3 & \textbf{4} & 0 & 7 & 0 &  & 27 \\
E2 & 47 mins &  & 27 & \textbf{102} & 84 & 213 & 29 &  & 4 & \textbf{5} & 3 & 12 & 1 &  & 24 \\
E3 & 60 mins &  & 136 & \textbf{281} & 0 & 417 & 70 &  & \textbf{17} & 12 & 0 & 29 & 2 &  & 20 \vspace{2pt}\\
$\mu$$\pm$$\sigma$ & 59$\pm$9 mins & & 81$\pm$45 & 179$\pm$75 & 28$\pm$40 & 288$\pm$91 & 57$\pm$20 & & 8$\pm$6 & 7$\pm$4 & 1$\pm$1 & 16$\pm$9 & 16$\pm$1 & & 24$\pm$1 \\
\midrule
C1 & 32 mins &  & 48 & \textbf{223} & 35 & 306 & 93 &  & 2 & \textbf{8} & 0 & 10 & 4 &  & 8 \\
C2 & 15 mins &  & 56 & \textbf{95} & 0 & 151 & 33 &  & 5 & \textbf{7} & 0 & 12 & 0 &  & 2 \\
C3 & 16 mins &  & \textbf{151} & 107 & 0 & 258 & 17 &  & \textbf{9} & 2 & 0 & 11 & 0 &  & 3 \\
C4 & 27 mins &  & 38 & \textbf{265} & 0 & 303 & 168 &  & 2 & \textbf{3} & 0 & 5 & 1 &  & 4 \\
C5 & 22 mins &  & 32 & \textbf{119} & 0 & 151 & 28 &  & 4 & \textbf{13} & 0 & 17 & 1 &  & 5 \\
C6 & 29 mins &  & \textbf{134} & 111 & 0 & 245 & 0 &  & \textbf{9} & 4 & 0 & 13 & 0 &  & 6 \\
C7 & 28 mins &  & 89 & \textbf{338} & 33 & 460 & 125 &  & 9 & \textbf{14} & 5 & 28 & 0 &  & 5 \\
C8 & 35 mins &  & 74 & \textbf{490} & 76 & 640 & 141 &  & 5 & \textbf{16} & 1 & 22 & 1 &  & 4 \vspace{2pt}\\
$\mu$$\pm$$\sigma$ & 26$\pm$7 mins & & 78$\pm$41 & 219$\pm$132 & 18$\pm$26 & 314$\pm$154 & 76$\pm$60 & & 6$\pm$3 & 8$\pm$5 & 1$\pm$2 & 15$\pm$7 & 1$\pm$1 & & 5$\pm$2 \\
\bottomrule
\end{tabular}
}
\caption{Time and interaction counts for participants in both user studies. Preview represents hovering to preview a version or alternative on the position, concept, or time palette. Reuse represents clicking on a version, position, concept, or time alternative to place it on the canvas. Generations represent text-to-image generations.}
\label{tab:designer_interactions}
\end{table*}

\section{User Studies of \system{}}
We conducted two user studies to evaluate how \system{} supports people in using a history of generated artifacts for reuse and communication in generative image composition workflows.
In summary, we aimed to assess:

\begin{itemize}
    \item[\textbf{RQ1}] \textbf{History Reuse.} How do people use \textit{new generations} vs. a \textit{history of generations} to create compositions in \system{}? 
    \item[\textbf{RQ2}] \textbf{Organization.} How do people use \textit{linear} (time palette) vs. \textit{non-linear} (concept and position palette) local alternatives? How do people use \textit{local} vs. \textit{global} alternatives?    
    \item[\textbf{RQ3}] \textbf{Collaboration.} Does \system{} support the use of project history for communication and collaboration?
\end{itemize}

\noindent To answer these questions, we first conducted \textbf{a study with creative professionals} who performed an open-ended image composition project with \system{}, and \textbf{a study with client collaborators} who used \system{} to work off an initial composition and already-populated alternatives. We address \textbf{RQ1} and \textbf{RQ2} by analyzing interactions and participants' descriptions of their experience in both studies. We address \textbf{RQ3} by collecting participant responses about their perceptions of sharing their history or using others' histories, and by comparing results across the two studies.

\subsection{Methods}

\noindent \textbf{Procedure.} We first asked participants demographic and background questions, then provided a 15-minute tutorial of \system{}. Participants then performed an editing task with \system{}. We instructed participants to think-aloud during their editing task while a researcher took notes. 
We then conducted semi-structured interviews covering their process, interactions with the system, benefits and drawbacks of \system{}, and desired improvements for \system{}.
Client collaborators also completed a final questionnaire with Likert item ratings asking about their concrete experience with select UI elements and interaction techniques (\eg "I thought [...] was useful.")
We recorded each study session and all participant interactions with \system{}. We conducted the studies remotely via Zoom and in person at a lab space depending on participant preference. Studies lasted 2.5 hours (\$40 gift card compensation) for creative professionals and 1.25 hours (\$20 gift card compensation) for client collaborators. \\ 

\noindent \textbf{Editing Task (Creative Professionals)}. Creative professionals performed a 60-minute open-ended editing activity designed to mimic a creative assignment. We provided professionals with the scenario \textit{``imagine you have been hired by a client to create a home surrounded by a landscape''} and an initial background image. 
We selected the task and background image (\ie a blue sky and green grass) to encourage open-ended exploration within the bounds of our system and generation limitations (\textit{e.g.}, works better for landscapes than portraits). \\

\noindent \textbf{Editing Task (Client Collaborators)}.
Clients performed a 30-minute editing task with the scenario \textit{``picture that you are a client who had hired a designer to complete a specific design scenario and had just received a copy of the designer's project in \system{}''} along with an initial composition and its complete history in \system{}. Each participant completed this task for one of three design scenarios that we assigned with a balanced random assignment.  
We then asked client collaborators to explore the project and edit the image to better fit their needs according to their assigned scenario.
The first author created all starting projects to represent ``designer'' projects in the client collaborator study. Each project took 2-3 hours to create due to generation speed (around 10 seconds per generation, limited by our hardware). \\

\noindent \textbf{Participants (Creative Professionals).} We recruited three creative professionals (E1-E3) from our professional networks. 
E1 had 13 years of experience as a graphic designer with regular clients in retail and e-commerce, E2 had 20 years of experience as a creative director with regular clients in diverse industries, and E3 had 8 years of experience as a graphic designer with regular clients in retail, fashion, and IT.
\changenote{
All professionals have used image generation tools (\eg Photoshop Generative Fill~\cite{adobephotoshop}, Midjourney~\cite{midjourney}, Dall-E~\cite{dalle}, Stable Diffusion~\cite{rombach2021highresolution}) before. On a 5-point scale (1=``never'' to 5=``always''), professionals reported using image generation tools ``sometimes'' (3/5; E2) or ``very often'' (4/5; E1, E3). On a 5-point proficiency scale (1=``very poor'' to 5=``very good''), they rated their proficiency as fair'' (3/5; E3), ``good'' (4/5; E1), and ``very good'' (5/5; E2). Thus, professionals were frequent and proficient users of image generation tools comparable to the one replicated in our system.
}
\\ 

\begin{figure*}[t]
    \centering
    \includegraphics[width=\linewidth]{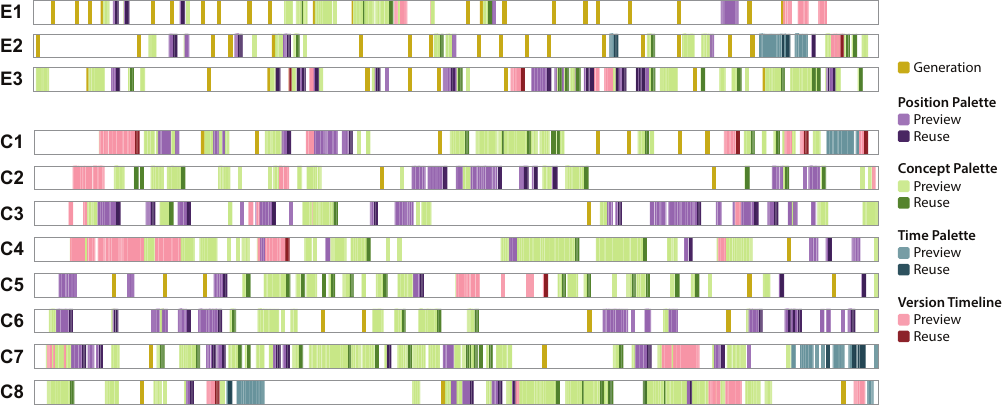}
    \caption{Creative professionals and client collaborators interactions with \system{} and the palettes. While creative professionals used the palettes more frequently as the project progressed, client collaborators first got acquainted with the projects by previewing via the version timeline.}
    \label{fig:interaction_log}
    \Description{
    An image displaying a horizontal bar chart with rows labeled E1, E2, E3 (top section) and C1 through C8 (bottom section). Each row consists of color-coded segments representing activities in a digital tool. The chart highlights patterns of interaction across different palettes and timelines. The legend on the right maps colors to each interaction.
    }
\end{figure*}

\noindent \textbf{Participants (Client Collaborators).} We recruited 8 client collaborator participants (C1-C8) from our professional networks who were familiar with generative AI tools for creative processes but did not need prior creative experience. 
\changenote{Client collaborators reported an average of ``fair'' image generation proficiency ($\mu=2.75, \sigma=1.29$) and use generative AI features ``sometimes'' ($\mu=2.75, \sigma=0.83$) for creative processes on the same 5-point Likert scales.}
\\ 

\noindent \textbf{Data Analysis.} We analyzed the user interaction logging data to identify the frequency of interactions across \system{} features. 
We used a German qualitative content analysis method~\cite{mayring2021qualitative} to analyze the transcribed interviews and observation notes we gained during the study. The analysis method is equivalent to thematic analysis~\cite{Braun2006,doi:10.1080/2159676X.2019.1628806}.
All authors thoroughly reviewed the transcription and observations individually by grouping, splitting, and merging codes to identify themes.
All authors then discussed their themes and collaboratively reorganized their codes until they agreed on common higher-level themes. We performed this analysis separately for each study, then combined common themes and discussed them using research questions.

\subsection{RQ1: Use of History in Image Generation}
All participants across both studies used \system{} to explore, preview, and reuse image generation alternatives from the prior project history (Table~\ref{tab:designer_interactions}). All participants stated in interviews that they appreciated the ability to use the history of the project in their editing task and would use \system{} in the future. \\ 

\noindent \textbf{How did creative professionals use history vs. generation?} 
Creative professionals started with blank projects, so they first generated images to compose their scene (four images were added to the history per prompt), then used the history more often as the project progressed (Figure~\ref{fig:interaction_log}). 
Creative professionals thus added to their composition more often with new generations ($\mu=24$ times; E1=27, E2=24, E3=20) than the history ($\mu=14$ times; E1=7, E2=10, E3=26) (Table~\ref{tab:designer_interactions}).
\changenote{
On average, creative professionals generated new content 61.2\% ($\sigma=20.4$) of the time and reused existing alternatives 38.8\% ($\sigma=20.4$) of the time (Table~\ref{tab:designer_interactions}).
When generating, all professionals iterated over their prompts at least once to refine generations and explored an average of 17 ($\sigma=1$) unique prompts per session, with an average of 3 ($\sigma=1$) semantic clusters in which they shifted prompts (\eg shifting from ``stone bridge'' to ``wooden bridge'').
Each semantic prompt cluster had an average of 6.2 ($\sigma=2.5$) prompts.
For example, E2 explored four variations for a flower bed (``blue wildflowers'', ``bouquet'', ``bush with flowers'', ``colorful flowers'').
}

\begin{figure*}[t]
    \centering
    \includegraphics[width=\textwidth]{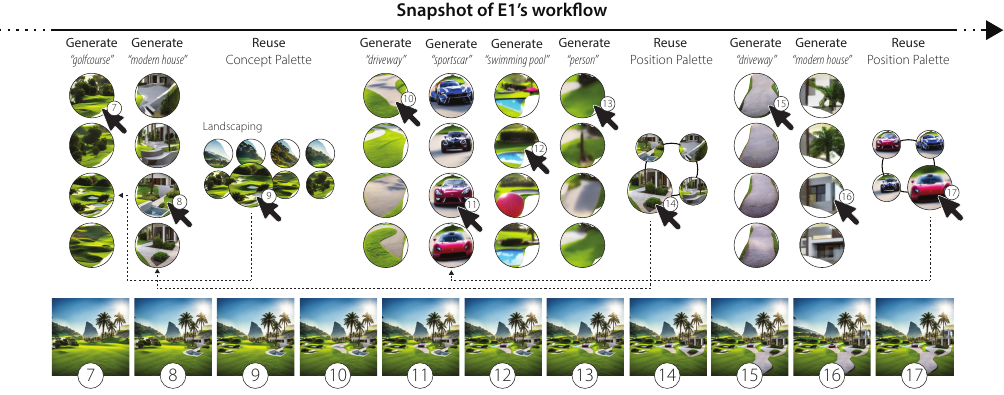}
    \caption{\changenote{Snapshot of E1's editing workflow (edit steps 7-17). E1 frequently switched between exploring ideas through generation (7, 8, 10, 11, 12, 13, 15, 16) and reusing past alternatives via the concept palette (9) and the position palette (14, 17) when composing the image.}}
    \label{fig:designer_1_workflow}
    \Description{A figure titled Snapshot of E1's workflow displays a horizontal timeline of an iterative image editing process labeled with steps 7 through 17. The diagram is organized into three rows showing the user's intent via text labels, the specific selection made from clusters of circular thumbnails, and the resulting composite image at the bottom. The workflow begins with steps 7 and 8, where the user generates a golf course and a modern house to establish the scene's background and main structure. At step 9, the user selects from a Reuse Concept Palette to refine the landscaping style. Steps 10 through 13 involve generating new elements including a driveway, a red sports car, a swimming pool, and a person. At step 14, the user employs a Reuse Position Palette to adjust the layout. Steps 15 and 16 show further generations for the driveway and modern house to refine their appearance. Finally, at step 17, the user accesses the Reuse Position Palette again to finalize the position of the sports car. The bottom row of images tracks the evolution of the scene from a simple landscape to a detailed final composition featuring a modern home with a pool, driveway, and car set against a tropical background.}
\end{figure*}

Creative professionals used the palettes more often towards the end of the session, hinting at the palettes' usefulness as the project progresses (Figure~\ref{fig:interaction_log}).
Only one creative professional used the time palette (E2). 
All creative professionals interleaved generation and history (Figure~\ref{fig:designer_1_workflow}) to perform their editing task (creating a home surrounded by a landscape) in different ways. We provide three vignettes:

\begin{figure}[!h]
    \centering
    \includegraphics[width=3.33in]{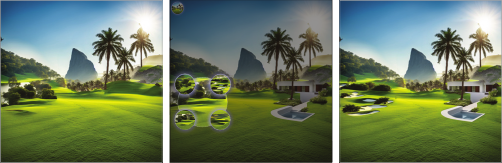}
    \caption{E1 reused a past design alternative of a golf course after making changes to the surrounding elements (here: generating a modern villa on the opposite side of the image).} 
    \Description{Three panels showing a workflow of E1. The first panel shows a lush green field with palm trees and a mountainous backdrop. The second panel shows circular overlays highlighting areas on the left side of the image, where design modifications are being considered while the right side of the image contains a modern house with pool. The third panel reflects changes applied to the left side of the landscape, including the addition of new design elements, while the rest of the scene remains consistent.}
    \label{fig:designer_1}
\end{figure}

E1 stated that they edited their image by starting from the outside (\textit{e.g.}, background scenery) then gradually \textit{``went with the flow of generations"} to create the main subject, a house (Figure~\ref{fig:designer_1}). E1 frequently previewed past alternatives between generations as they went to see which alternative fit best (Figure~\ref{fig:interaction_log}). E1 noted they previewed and used past alternatives for key parts of the image (\textit{e.g.}, golf course, and house), as those were the generations that \textit{``took [him] longer in the selection process because a few alternatives would have looked good"}.
For example, E1 was once between two ideas as they noticed that the modern house and palm trees they recently generated no longer fit with the casual style of a golf course they added earlier. To decide whether to revise the golf course or the house, E1 used \system{}'s position palette to view alternative golf courses. E1 liked one with a manicured look to match the modern house so they placed the golf course into the image.

\begin{figure}[!h]
    \centering
    \includegraphics[width=3.33in]{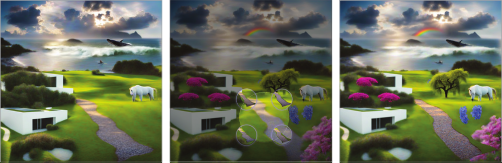}
    \caption{E2 generated a path to the house and spent the next 10 minutes adding more elements to the image such as a tree and flowers. E2 only now realized that they prefer a past design of their path.}
    \label{fig:designer_2}
    \Description{
    Three panels showing a workflow of E2. The first panel shows a modern house surrounded by greenery, a winding path leading to the house, an elephant on the lawn, and a dramatic sky with sunlight breaking through clouds. The second panel introduces circular overlays highlighting specific areas where changes are being considered. There are new elements in panel two compared to panel one, such as colorful flowers, a tree, and other elements around the path. The third panel shows the updated design, with a past version of the path that was preferred.
    }
\end{figure}

E2 first added a landscape to set the structure of the image (\textit{``where to put sea, sky, land?''}) then added a subject (house) and the surroundings (\textit{e.g.}, garden).
E2 used the palettes towards the end of their process \textit{``to see if there were any objects that were better''} (\textit{e.g.}, paths, plants).
For example, they picked a different path they decided against earlier, as it now looked best with their current version (Figure~\ref{fig:designer_2}).
E2 then added newly generated elements to the image, including flowers and trees, and realized how yet a different path might go better with the updated surroundings, so they updated the path with a prior path alternative.
Then they previewed and selected different flowers as they preferred the old flowers with the new path.

\begin{figure}[!h]
    \centering
    \includegraphics[width=3.33in]{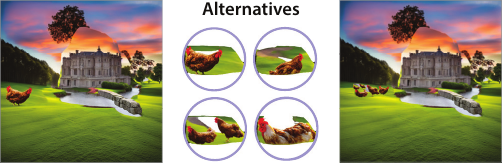}
    \caption{E3 used \system{} to reuse past designs (here: generations of chickens) and created alternative variations of them by resizing and moving them before pasting onto the image.}
    \label{fig:designer_3}
    \Description{Three panels depicting a workflow of E3. The first panel features a mansion with a grassy landscape, a pond, a tree, and one chicken placed next to the pond. The middle panel displays circular thumbnails labeled 'Alternatives,' showing different variations of chickens in various sizes and positions. The third panel shows the final design, where resized and repositioned chickens have been incorporated into the landscape.}
\end{figure}

E3 started by regenerating their background to create a beach scene and then reverted it to create a house because they wanted to keep the beach in the history to potentially revisit this design choice later. 
E3 noted that it was \textit{``hard to imagine what I want''} so they liked the idea of creating the project first and then playing with their past designs and recombining them.
E3 frequently used live preview (487 times, Table~\ref{tab:designer_interactions}) to assess when any past designs might fit better according to their personal preference and then added preferred designs.
At the end of the session, E3 decided to \textit{``move it all to the left''} increasing exploration as they attempted to rebuild the scene by reusing and recombining prior designs. \\

\noindent \textbf{How did client collaborators use history vs. generation?} 
Client collaborators started with a project with a pre-populated history, and thus instead explored the history first and generated images when their desired outcome did not already exist. Therefore, clients used alternatives from the history ($\mu=12$, $\sigma=6$) more often than they used new generations ($\mu=5$, $\sigma=2$).
\changenote{On average, 75.6\% ($\sigma=10.2$) of clients' edits were reusing alternatives and 24.2\% ($\sigma=10.2$) new generations (Table~\ref{tab:designer_interactions}).}
All client collaborators used the position and concept palettes while only 3 of them (C1, C7, C8) used the time palette (Figure~\ref{fig:interaction_log}).

Client collaborators first explored the history of alternatives using hover to preview (Figure~\ref{fig:interaction_log}), then added a prior alternative to the image when they disagreed with the designer's choice or wanted to change the scene to match a higher-level goal (\eg C8 recombined past designs to get a darker setting). 
Participants described that they perceived past designs as starting points (C3, C4, C5) that give options (\eg \textit{``I want to see if there are other nice houses?"}) when picking up someone's idea to improve on it. For example, C6 said \textit{``it looked better in my eyes"} after explaining why they reused a designer's past house design. 
Client collaborators all changed regions of the composition and many made large changes (6 changed the background, 6 changed the main image) (Figure~\ref{fig:client_outcomes}). 
C6 uniquely placed a past design (a barn) in the middle of the scene to explore how the scene would look with a house in the middle, then generated a new house over the barn after they were satisfied with the placement.
After bringing past designs back into the image, client collaborators frequently applied transformations to them (i.e. move, resize, and rotate) to make them fit better into the current state of the image. 
Client collaborators then most often used generation to add concepts that the designer did not explore (\eg a ``horse'', or ``grass'' instead of existing landscaping options) rather than for generating new versions of previously explored concepts. \\

\noindent \textbf{History reuse via preview over addition.}
In both studies, participants used prior design alternatives from their history in their image (from 5 to 26 total alternatives used in the final image, Table~\ref{tab:designer_interactions}), but participants experimented with alternatives by hovering to preview them hundreds of times during a project (from 179-781 hover actions per participant).
All participants described that the most important interaction with the system was to explore past designs by hovering on palette items to preview them in the image.
This is also supported by participants' interactions with \system{} (Figure~\ref{fig:interaction_log}).
Participants highlighted the importance of being able to preview past designs \textit{``next to the alternative''} (C7), which made it \textit{``easier to visualize"} (C8) and \textit{``to directly see what it would look like"} (C7) \\

\noindent \textbf{History reuse via paste over rasterize.} For both studies, participants often applied transformations to alternatives when they brought them back (\ie move, resize, and rotate) to fit better into the current state of the image. To add alternatives back 10 of 12 participants used paste more often than rasterize. 
Participants used paste when the alternative matched the background, and typically used the time-consuming rasterize option more sparingly to smooth out artifacts from alternative/background mismatches. \\

\noindent \textbf{Creative professionals valued history in generative workflows for control and iterative harmony improvement.}
Professionals favorably compared \system{} to their prior workflows with generative AI due to the control provided by the history. E1 mentioned that in a prior project, they wanted to generate ``a pair of skis'' similar to a pair they had generated before but lost access to.
After long back and forth they had to give up as their existing tool would not provide them with the desired outcome of the generation, but noted that they would be able to bring back the prior version in \system{}. 
On a similar note, E2 mentioned that \system{} was more ``controllable'' than Photoshop. 

All creative professionals worked iteratively with the palettes provided by \system{} to achieve harmony in their scenes.
For instance, E1 described when exploring prior house alternatives \textit{``this house looks better in the scenery which I just realized after all my changes''}.
E1 mentioned that the harmony between previously generated objects in the scene changes whenever new objects are added to the scene.
They describe it as ``object dependency'' which is affected by new generations that are added to the image.
Similarly, E3 noted how \textit{``seeing it [objects] in the surroundings led me to taking a different turtle''} which demonstrates how designers make their next design decisions by trying to achieve harmony between elements in their image.
E2 also noticed how later in their process they suddenly liked a different flower better when previewing it, suggesting the preview makes this evaluation of harmony easier for designers. \\ 

\noindent \textbf{Client collaborators valued history for quality, efficiency, and learning. }
All Client collaborators used the history more than generation and preferred \system{} when asked to compare it to traditional methods like the Generation View in our interface or the Photoshop Generative Fill feature. Clients reported they preferred to use the history compared to generation as it was easier to reuse past designs rather than generate new designs as they are unfamiliar with prompting (C8), worse at prompting than the designer (C5), that they already trust the designer to make good alternatives (C2, C4), and that they find it hard to put their thoughts into words (C2). 
Participants noted that seeing the designer's prompts with the prompt view option helped them articulate their own prompts (\eg C2 mentioned the prompts provided helpful orientation and C4 mentioned they would not have thought about some prompts). \\


\noindent \textbf{History reuse supports varied collections.} E3, C2, C5, and C6 used discarded alternatives to create varied collections including a flock of chickens (E3) and a village of houses (C6). E3 prompted for a chicken, received four generated alternatives, and selected one to add to the scene. E3 later decided they wanted to create a flock of chickens and thus revisited the history to add the other discarded chickens with a variety of poses and appearances to the scene. C2 similarly created their village through the reuse of transformed versions of past house designs and discarded alternatives, letting them create variations while keeping consistency.

\subsection{RQ2: Organization of Alternatives}

\begin{figure}[t]
    \centering
    \includegraphics[width=3.33in]{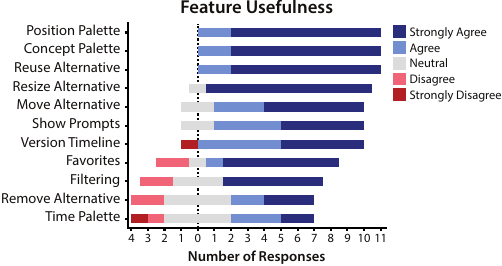}
    \caption{Participant ratings on the usefulness of the features of \system{}. Participants rated the positional and concept palettes as more useful than the time palette.} 
    \label{fig:likert}
    \Description{A horizontal bar chart titled 'Feature Usefulness' displaying participant ratings of features in HistoryPalette. Features are listed on the vertical axis, including Position Palette, Concept Palette, Reuse Alternative, Resize Alternative, Move Alternative, Show Prompts, Version Timeline, Favorites, Filtering, Remove Alternative, and Time Palette. The horizontal axis represents the number of responses, with bars color-coded by agreement level: dark blue (Strongly Agree), blue (Agree), gray (Neutral), light red (Disagree), and dark red (Strongly Disagree). The Position Palette and Concept Palette received the highest 'Strongly Agree' ratings, while the Time Palette received the lowest ratings, showing less perceived usefulness.}
\end{figure}

\noindent \textbf{Participants used and preferred position and concept palettes over the time palette.}
All participants used position or concept palettes (non-linear, new) more often than they used the time palette (linear, existing) (Table~\ref{tab:designer_interactions}). 
10 of 12 hovered on the concept palette most often to preview alternatives (C3 and C6 hovered on the position palette most often), and 9 of 12 clicked on the concept palette most often to bring an alternative back (E3, C3, and C6 clicked on the position palette most often). 
4 of 12 hovered on the position palette most often and 3 of 12 clicked on the position palette most often to bring an alternative back.
7 of 12 participants never used the time palette.
In Likert scale ratings client collaborators indicated that the position palette was most useful followed by the concept palette, while the time palette was rated the lowest out of all system features (Figure~\ref{fig:likert}). \\

\noindent \textbf{Concept and position palettes reduce the search space.} An advantage of concept and position palettes is that they reduce the search space given a user goal. Creative professionals produced 75-95 alternatives and 62-72 alternatives post-filtering (removed ~20\% of alternatives). Creative professionals had 6 ($\sigma=1$) concepts with 11 ($\sigma=5$) alternatives, and pixel positions with 10 ($\sigma=3$) alternatives each. Creative professionals thus experienced up to a 11x (concept palette) or 7x (position palette) reduction in the search space (similar reductions for clients). Participants used nearby alternatives in the position and concept palette to flicker between a small set of position or concept alternatives, and skimmed concept palette labels to identify topics of interest (Figure~\ref{fig:interaction_log}). \\

\noindent  \textbf{History palettes support flow. } E1 and E3 frequently used the position and concept palettes to preview alternatives, and reported that \system{} thus let them stay in the flow as they worked on their design. E1 shared that using \system{} is \textit{``like a river and you flow with the program''} as previewing and using alternatives did not distract their attention from the final image. E3 described:
\begin{quote}
    \textit{``I love that I can just go with the flow in creative process here [...] you can just click through and see what could be better or not, it's just a flow.''\\}
\end{quote} 


\noindent \textbf{Participants previewed but rarely reused global alternatives.}
Participants often hovered over prior versions to preview them but rarely clicked prior versions to bring them back. 
Client collaborators specifically reported that they used the prior whole project versions to understand the overall process to build the project. A few clients restored full prior versions but still used alternatives from a later version. For example, C5 selected an earlier version of the image with fewer elements and then used recent alternatives to build the scene.

\subsection{RQ3: Use of History for Collaboration}

\noindent \textbf{Experts valued sharing own history for collaboration.} E1, E2, and E3 stated \system{} would support them in communicating with others. 
All participants would use \system{} to create multiple versions to communicate with clients, but participants varied in whether they thought it would be appropriate to give the tool directly to the client. 

E1 noted \system{} would be particularly helpful in supporting \textit{``interactive design exchange with the client''} because it \textit{``allows design finding within 15 seconds.''} E1 also wanted to give the entire interface to a client because it is easy to use for clients and would allow clients to make small adjustments --- in fact, the company they worked at attempted to make a custom internal tool for a similar purpose. 
However, they mentioned they could see some designers not wanting clients to be able to modify the final design with the tool (\textit{e.g.}, to keep a consistent style like photographers who do not share the raw image). 
E3 also expressed that clients may not be interested in seeing the final project, but that it could be useful for designers to walk clients through alternatives they considered in the design (\textit{e.g.}, \textit{``we decided to go for a house but we could have also gone with a bunker instead''}).
E2 noted that they would not want to \textit{``show 100 different options to the client''} but they may want to select alternatives for the client to see (\textit{e.g.}, a client could select between two different houses).

E2 and E3 valued history sharing to collaborate with other designers. 
E2 expressed \textit{``if a different designer has to open the files, it's very useful!''} She elaborated that if a client gave feedback on specific flowers in the final image, another designer could turn on prompt history to see what prompt was used to create those flowers, then change or improve the prompt (\textit{e.g.}., using the position palette). 
E3 noted how \system{} would be useful in collaborative settings where \textit{``one lives off of ideas from others"} (\textit{e.g., ``let's take my mountain and your house"}). \\

\noindent \textbf{Client collaborators valued the ability to understand and communicate about the designer's process. } 
Participants perceived \system{} as a way that allows them to go back and not only see the history of the project but also understand why designers' decisions were made.
C4 described the ability to see the history as getting to talk to the designer. 

Participants highlighted how \system{} allows exploring past designs as a means of getting a better understanding of the designers' process and reasoning.
Participants understood the palette of past designs to show the design space that the designer's past explorations covered.
Participants noted that they could use \system{} find spaces where designers already explored alternatives that were not good (C6), left room for exploration (C8), and potentially communicate the gap (C4). C4 stated they may ask \textit{``Why did you not change the house that often?''}
Participants reported that viewing the prompts further provided them with a way to understand the designer's reasoning including their goal (C7), content of focus (C7) and intentions (C6). 

Participants finally noted that \system{} allows clients to take a step towards the designer role. C4 specifically mentions that \system{} allows them to \textit{``change something and then go back to the designer and tell them how [they] like it"}.

\subsection{Future Use and Improvements}


\noindent \textbf{Creative professionals wanted to use HistoryPalette for long and open-ended projects. } Creative practitioners E1 and E3 suggested \system{} would be particularly helpful for large and long projects. E3 imagines that the system would be particularly helpful when working on big projects with 200 objects or more as \textit{``you can go through them all and have them collected there''} or easily change an object even late in the project (\textit{e.g.}, clicking on a plant to see what else they did there).
E1 described that \system{} solved a common problem in their usual process: 
\begin{quote}
\textit{``If I lose a small object and only realize after 1.5 hours that I lost an object that I cannot really bring back [...] I need to push CTRL + Z all the time just to bring it back, here I can always just bring it back. [...] I have this problem very often, it's among my top 3 problems.''}
\end{quote}

All participants stated the tool would be strongest for open-ended settings. E1 described that they know 70\% of what they want to create ahead of time, but 30\% comes from going with the flow such that \system{} would be strong for the open-ended 30\%. E2 reported the tool would be most useful when they were considering a range of possible directions. \\

\noindent \textbf{Improving controllability of \system{}}. Participants raised areas to improve \system{}. Participants noticed artifacts from placing an earlier generation on a new background (\textit{e.g.}, mismatched edges), and thus reported that they would like the option to reuse the foreground object of a prior alternative rather the entire mask. 
E3 also mentioned that the position palette was helpful but could be difficult to control access to prior small generations (\textit{e.g.}, the mask of a lamp post). While E3 used the concept palette to find the lamp post, they reported that zoom would be useful in the future. 

\begin{figure}[t]
    \centering
    \includegraphics[width=3.33in]{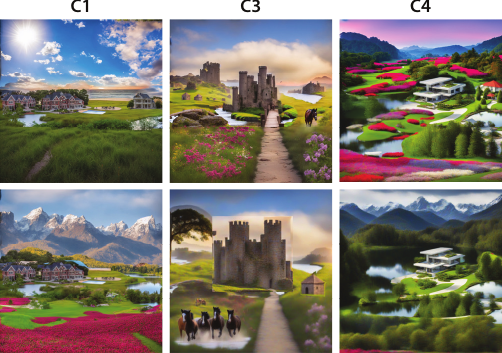}
    \caption{Client's original (top) and edited (bottom) images. C1 and C4 both changed major background elements, while C3 changed the scale of the castle and generated new horses.}
    \label{fig:client_outcomes}
    \Description{Two rows of images showing the original (top row) and final (bottom row) edits made by clients (C1, C3, and C4). For C1, the original image features a village near a grassy field under a blue sky, while the final edit adds vibrant red flowers in the foreground and snowy mountains in the background. For C3, the original image shows a castle on a hill with a pathway and flowers, while the final edit enlarges the castle and adds a group of horses on the path. For C4, the original image depicts a modern house surrounded in a hilly environment with red flower patterns across the landscape, while the final edit depicts a modern house surrounded by green hills without the red flower patterns and a changed mountain landscape in the background.}
\end{figure}

\section{Discussion}
We reflect on the creative professional and client studies to summarize our key findings and reflect on our design goals (\textbf{G1-G4}), then share limitations and opportunities for future work. 

\subsection{Study Reflections}

\noindent \textbf{G1: Palettes of Alternatives for Discovery.} Creative professionals and collaborators frequently used unique features of \system{} (position palette, concept palette) to discover and then reuse alternatives. 
Participants rarely used or preferred the time palette that most closely mirrors existing support for partial-project alternatives (\textit{e.g.}, layers created by Photoshop Generative Fill~\cite{adobephotoshop}).
Across both studies, participants used the concept palette most often, followed by the position palette and version timeline (but they rarely clicked the version timeline to revert). 
We were initially surprised that the concept palette was used more often than the position palette, as participants expressed the most enthusiasm when talking about the position palette for discovering object alternatives. In practice, the concept palette was always open and easy to browse for alternatives of interest (\textbf{G1}). 

\textbf{G2: Support rapid reuse of alternatives.} Across both studies participants shared that reusing alternatives was faster and more predictable than image generation such that they would prefer to reuse their palettes when possible. In both studies, participants successfully added alternatives into their image via rasterize or paste (from 5 to 26 total alternatives used in the final image, Table~\ref{tab:designer_interactions}), but participants experimented with alternatives by hovering over them in the palette hundreds of times during a project (from 179-781 hover actions per participant). \textbf{G2} may be best supported by the combination of right-at-hand alternatives (\textbf{G1}) with hover functionality that enables rapid experimentation of design alternatives. 

\textbf{G3: Provide history filters and favorites. } 
Participants found histories to be more useful as they got longer which emphasizes the need for history management features. 
While participants had the option to toggle our generation alternatives filter on or off, all participants kept the filter on. Our filter culled alternatives but participants never mentioned the filter positively or negatively. Participants did not use the favorites feature during the limited editing time, but creative professionals noted that this feature may be useful for saving their own alternatives in collaboration (\textbf{G4}). 

\textbf{G4: Support multiple audiences.} Both sets of participants expressed excitement about using the system for collaboration (\textit{e.g.}, to help another designer see the prompts, or learn about designers' prompts), indicating the potential for collaborating using non-linear edit histories (\textbf{G4}). However, client collaborators and creative professionals had occasionally mismatched expectations about how much of the design should be editable vs. stuck in place (\textbf{G4}).

\subsection{\changenote{Limitations}}
\changenote{
Our evaluation prioritized deep, qualitative insights with expert professionals (N=3) over broad generalization.
However, we acknowledge that our findings are specific to experienced graphic designers working on unconstrained image editing tasks (e.g., landscapes).
While we found promising insights on the usefulness of \system{} in bottom-up image generation workflows, we think conducting further validation with a larger and broader group of participants is needed to further confirm our findings.
For example, our findings may not extend to creative domains with stricter functional constraints, such as architecture or UI/UX design, where physical laws or design guidelines limit the variability of generations.
Similarly, novice designers might populate palettes with trial-and-error experiments rather than the reusable alternatives that were generated by experts.
}

\changenote{
We specifically targeted ``bottom-up, material-centric'' design workflows~\cite{sterman2022CreativeVersionControl} (\eg~image compositing, creating collages). However, we currently do not support top-down workflows where the user has a clear end goal in mind and works in a linear progression towards a single measurably better result (\eg~creating high-fidelity images from iteratively refining prompts, building complex pipelines in comfyUI~\cite{comfyanonymous2025comfyanonymous}). Future work could also support history exploration and reuse for top-down workflows by tracking changes in model inputs (\eg prompt or pipeline changes) instead of in outputs (\eg image generations). For example, an LLM could be used to identify a prompt change and annotate it with a label to be included in the history palettes. Users could then reuse parts of past prompts by letting the system prompt an LLM to integrate a past partial prompt alternative into the current prompt for the current canvas. Furthermore, because prompts are often influenced by prior prompts in iterative prompt refinement workflows, Fuzzy Linkography~\cite{smith2025fuzzy} could be used to compute and visualize the semantic links between prompt alternatives.
}

A technical limitation is bringing objects back to new locations without introducing quality artifacts due to obvious boundaries between old and new object backgrounds. To mitigate this issue, participants often selected locations for the object with a similar background, accepted the collage-like feel, or used rasterize to smooth out artifacts.
While rasterization reduces boundary artifacts, it changes small details the image composition.
Better object insertion models~\cite{winter2024objectdrop, winter2025objectmate, ruiz2025magic} and extracting alternatives from their backgrounds could mitigate boundary artifacts when reusing alternatives.
The creative professional who had significant experience with image generation reused most alternatives without introducing artifacts. As users may be able to improve their ability to use the interface and image generation models (\eg Nano Banana Pro~\cite{nanobanana}, FLUX.2~\cite{flux}, Seedream 4.0~\cite{seedream2025seedream40nextgenerationmultimodal}) are rapidly improving, we expect this issue to diminish over time.

\subsection{Implications and Future Work}

\noindent \textbf{Implications for prompt-based workflows.}
Creators deal with uncertainty in prompt-based workflows as text-prompts are often underspecified when generating high-fidelity images, but each new generation costs money and takes time. Prior work addresses uncertainty by supporting creators refine their prompts ~\cite{brade2023promptify,wang2024promptcharm,feng2023promptmagician} or visualizing potential prompt outputs~\cite{almeda2024prompting}.
\changenote{
Our results showed that access to on-demand alternatives (\textbf{G1}, \textbf{G2}) promoted reuse rather than generation for both user groups.
Reducing uncertainty by revisiting the palettes improved professionals' flow and increased client's confidence in achieving high-quality results with limited prompting experience.
While generations in our system took slightly longer than those of traditional tools (\textasciitilde30s vs. \textasciitilde15-25s), our results suggest that reuse was mainly favored due to its predictability (knowing exactly what will be added) rather than just to avoid delays. While clients primarily avoided generation due to uncertainty (reusing alternatives for 75.6\% of edits), designers (who reused 38.8\% of the time) may have been more impacted by delays and may have done even more trial-and-error if generation were instant. Overall, while faster future models allow more rapid iteration, we expect the preference for re-use to persist because of a faster accumulation of reusable alternatives.
}

While creators and clients made use of the palettes to experiment with and use alternatives, we do not know what the long-term impact of such a tool might be. 
If using prior alternatives is significantly easier, cheaper, or faster than generation, users may slowly converge to a set of alternatives and encounter design fixation~\cite{cross2024expertiseindesign}. Alternatively, users may try new compositions with ease of palette use (as easy access to history can de-risk exploration~\cite{shneiderman2007creativity}).  Future work may explore how providing easy to access alternatives will impact the diversity of generations over time. Further, as professionals found \system{} to be more useful as the project progressed, it may be particularly beneficial for professionals to reuse palettes across projects. To avoid self-plagiarism, \system{} could indicate or fully remove alternatives creators previously used in a final composition for prior projects (\ie to ``retire'' the alternative).
\\


\noindent \textbf{Implications for sharing project history with others.} While prior work primarily explored project histories for personal reuse~\cite{Rawn2023,dang2023worldsmith,Angert2023SpellburstAN} or sharing a creator's process~\cite{klemmer2002web,grossman2010chronicle}, \system{} uniquely explores providing collaborators with another person's project history to directly explore and reuse alternatives or provide feedback. 
\system{} provided easy history exploration by matching client goals (\textit{e.g.}, preview alternative houses by clicking on the house or looking for houses) and low effort previews to quickly explore alternative designs (\textit{e.g.}, compared to the time and effort for new generations, or time-based history alone). Our client study demonstrated that clients without experience with text-to-image generation could use the history to create new compositions (Figure~\ref{fig:client_outcomes}). 

Prior work also explores using AI to create artifacts that ground conversation between creators and their collaborators~\cite{ko2022we,chung2023artinter,wang2024roomdreaming} --- \textit{e.g.}, by creating moodboards to establish project direction~\cite{chung2023artinter}, synthesizing references to inform webtoon design~\cite{ko2022we}, or planning initial room designs~\cite{wang2024roomdreaming}. Such systems address early stages of the design process when communicating shared direction~\cite{chung2023artinter} and achieving a breadth of exploration may be the goal~\cite{wang2024roomdreaming}, but \system{} demonstrates an opportunity to use the project history to provide feedback in later stages of the design process. In the later stage, collaborators may no longer establish direction but rather provide grounded feedback by previewing alternatives, understand designer constraints for the current choice (\textit{e.g.}, the alternatives do not look better), or even use the palettes to directly tweak the final design within a constrained design space.

As our studies demonstrated the promise of using palettes of alternatives within a single project for personal or collaborative use, an opportunity exists to explore how to manage, share, and reuse palettes across multiple projects or with multiple collaborators. For example, if a creator is working on a project with a colleague, they may combine their palettes. Or, a content creator may sell palettes to people hoping to emulate their style (\textit{e.g.}, like video color grading or photography editing packs).
A creator may keep a stack of palettes for different types of projects. Future work may explore how to suggest new palette additions (\textit{e.g.}, full palettes or single alternatives) based on the user's current palettes or projects.\\

\textbf{Generalization to direct manipulation workflows. }
\changenote{
\system{} supports image composition via generative in-painting, a core task of digital artists creating AI-supported art in tools like Photoshop Generative Fill~\cite{adobephotoshop}.
However, users sometimes also use direct manipulation tools (\eg brushes, retouching) alongside generation.
For example, an artist may use a brush tool to paint a new rock in a generated river, or add a highlight to a generated rock.
While direct manipulation tools can be trivially integrated into \system{}'s image editor, interactions with direct manipulation tools would currently not be recorded in the history palettes.
To support the reuse and exploration of direct manipulation alternatives, future work should explore how to extend our palettes to handle direct manipulation.
However, integrating direct manipulation edits as reusable alternatives in our palettes comes with unique challenges: segmentation (how to determine where an ``alternative'' begins and ends in a noisy stream of interactions), semantic labeling (how to label the alternative to be indexed in our palettes), and re-use (how to re-apply the alternative).
To address these challenges and support history exploration and reuse in workflows with direct manipulation, we propose extending \system{} with automatic segmentation and labeling of interactions, similar to prior work on workflow capturing~\cite{grossman2010chronicle, Pongnumkul2011PauseandplayAL, Truong2021AutomaticGO, Son2025ClearFairyCC}
In the future, we can use object recognition to identify new objects added via direct manipulation (e.g., ``rock''), then give the objects descriptive labels with a large multimodal model (``gray rock with painterly strokes''), and trivially add these alternatives to the existing palettes.
However, this will not work for abstract direct manipulations (\eg strokes) because they do not represent isolated, semantically meaningful alternatives.
Thus, we explore how to automatically segment, label, and reapply abstract direct manipulation interactions in the future.
Prior work explored using heuristics such as tool switches~\cite{grossman2010chronicle}, pauses~\cite{Pongnumkul2011PauseandplayAL}, or visual screen changes~\cite{Truong2021AutomaticGO} to segment based on recorded interactions. Future work could explore using similar heuristics (\eg pauses between strokes) to segment abstract direct manipulations into meaningful alternatives.
Once interactions are segmented into alternatives, Vision Language Models (VLMs) offer a promising solution for automated labeling as of their impressive capabilities in describing visual differences~\cite{Peng2022} and inferring user rationales from screen recordings~\cite{shaikh2025creating, Son2025ClearFairyCC}. For example, a VLM could observe the canvas state before and after a segment of consecutive interactions (\eg~a set of brush strokes) to generate a label (\eg~``tree''), or use verbalized intentions by the artist to generate an ``intention label'' (\eg apply a brush stroke highlight to match background lighting).
We could then allow reuse of a segmented and labeled alternative by creating a reusable macro that will re-apply its group of direct manipulation interactions, inspired by prior work in Computer Graphics that created reusable macros from artist demonstrations~\cite{grabler2009generating}.
\\
}

\noindent \textbf{Scaling to longer histories.}
\changenote{
\system{} builds upon workflows in existing commercial tools and expands upon existing history keeping mechanisms (\eg, chronologically ordered generations) by providing users access via position or concept to support discovering past design alternatives (\textbf{G1}).
For the scale of our professional sessions ($M=59$ mins, $STD=9$, 75-95 total alternatives) where professionals had an average of 10 alternatives per position palette, our palettes alone reduced the search space by 11x (Concept Palette) and 7x (Position Palette) compared to linear logs. Thus, professionals primarily used the concept and position palettes rather than our time palette (chronological order) for reuse.
However, as histories scale to hundreds or even thousands of alternatives, potentially across multiple sessions, visual clutter (for position palettes), and excessive scrolling (for concept and time palettes) may pose challenges. We see promising future directions for applying approaches for managing scale in generation histories with mechanisms through filtering, ranking, searching, and hierarchical grouping of alternatives.
Filtering limits the visible alternatives based on specific criteria to reduce clutter. For example, the position palette currently shows all alternatives registered for a location, which may include the foreground object (potted plant) and background objects (porch, landscape). In the future, we can allow users to select an active object and apply a context-aware filter to the palette to only show alternatives that are semantically related to that object (\eg only other potted plants).
To reduce scrolling in the palettes, we can rank alternatives (\eg based on frequency) to show more relevant ones first. For example, in the concept palette, frequency-based ranking could show users their most frequently used and thus potentially most useful alternatives, which could also serve as an automatic way to label alternatives as favorites (\textbf{G3}).
Our palettes can be trivially extended with a search tool to quickly find specific alternatives in largely populated palettes. For example, a user could search for a ``blue striped rug'' instead of having to scroll through hundreds of alternatives in the palettes.
Finally, alternatives can be further organized via hierarchical groupings. In the future, we will explore grouping together related alternatives (\eg based on color, style, size) into categories that can be expanded to access their alternatives. For example, the position palette could have a single node for a category (\eg ``animals'') that a user can expand via click to access its underlying alternatives (\eg ``horse'', ``dog'', etc.), similar to concept grouping in the concept palette.
\\
}

\noindent \textbf{Dependency-aware reuse of alternatives. } \changenote{
Creators iteratively branch, discard, and refine ideas in their creative workflows, creating a graph of exploration paths (many of which abandoned).
\system{} augments existing workflows by providing users with palettes to access and reuse all explored alternatives within their history.
Currently, we treat each creative step (\eg an inpainting action) as an independent, standalone alternative and fill our history palettes with one item for each alternative.
Branching between exploration paths through re-use from our palettes thus mirrors existing practice in commercial tools where users manually copy-paste elements or layers between different file versions.
However, a sequence of individual steps often shows that there are dependencies between the steps. For example, a higher-level creative goal (\eg adding a house to an image) is typically completed through several dependent alternatives in a row (\eg first adding the walls, then the roof, and then the windows). Individually reusing just the ``roof'' step often makes no sense if the ``walls'' step was not applied yet. 
Thus, sequences of dependent alternatives form semantically coupled sequences (or sub-trees in a history graph).
Therefore, an exciting direction for future work is to respect these dependencies between alternatives by letting users reuse whole sequences (or sub-trees) instead of single alternatives when switching and merging between branches.
Although not supported by existing commercial tools, prior work explored supported switching and merging between multiple explored paths, often in parallel~\cite{terry2004variation}, by respecting dependencies between versions~\cite{zuend2017storyversioncontrol, terry2004variation, dang2023worldsmith}.
In the future, we will explore populating our palettes with items for whole reusable sequences instead of only items for single alternatives. 
However, one challenge will be identifying these semantic subtrees in users' history graphs.
One way to identify and apply such subtrees is using methods like Tree Edit Distance~\cite{PAWLIK2016157}, similar to how Zünd et al.~\cite{zuend2017storyversioncontrol} enables merging between two distinct history trees of collaborators. In addition, LLMs to find subtrees by segmenting sequences of prompts.
Providing users with access and reuse of subtrees through the palettes then supports easier switching and merging between exploration paths within their history trees.
}

\section{Conclusion}
We introduced \system{}, a system for exploring and reusing prior alternatives to support image composition via generative in-painting. \system{} features four history views: a position palette that lets creators access alternatives at the current position, a concept palette that lets creators access alternatives clustered by prompt themes, a time palette that lets creators access alternatives by time, and a linear version timeline that lets creators access the version history of the entire project.
Our user studies demonstrated how semantic organization of generative histories can benefit both creators and their collaborators.
Across both studies, all participants used our semantic organization methods (position and concept palette) more often than the traditional time-based organization (time palette).
While creative professionals used prior alternatives to inspire new ideas and test alternatives, client collaborators used \system{}'s prior generations ($\mu=11.88$ times, $\sigma=6.09$ times) more often than new generations (4.62 times, SD=1.73 times) to edit the image because they preferred the efficiency and control of reusing prior generations.
All participants also expressed enthusiasm about using \system{} for collaborative work between creatives and clients (\eg clients providing feedback, learning from expert prompt strategies, and exploring personal preferences).
We hope our work inspires future research into improved history-keeping mechanisms for generative workflows to enable easy access and reuse of prior design alternatives, and communication between collaborators through shared histories.


\bibliographystyle{ACM-Reference-Format}
\bibliography{references}

\appendix
\newpage
\section{Appendix}
\label{appendix}

\subsection{Examples of our filtering results}
\begin{figure}[t]
    \centering
    \includegraphics[width=2.5in]{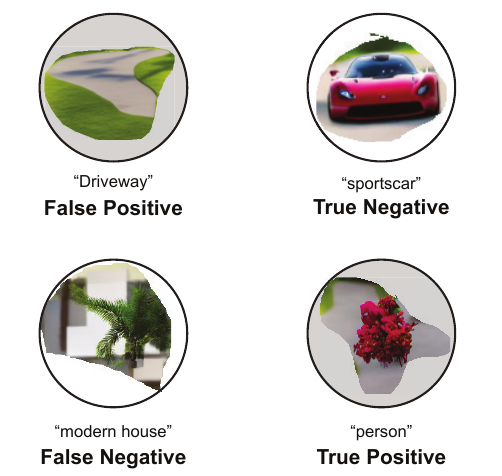}
    \caption{Example successes and errors of our filter that removes incorrect generations. The grey background indicates the system predicted a generation error. The system incorrectly classifies the driveway as a generation error (False Positive), correctly classifies the sportscar as a correct generation (True Negative), incorrectly classifies the plant generated for ``modern house'' as a correct generation (False Negative), and correctly classifies the flowers generated for the prompt ``person'' as a generation error (True Positive).}
    \label{fig:filter_eval}
    \Description{
    An image showing four circular thumbnails illustrating examples of successes and errors made by a filtering system for identifying incorrect generations. Each thumbnail is labeled with a prompt and classification result: Top left: A curved driveway labeled 'Driveway' with the classification 'False Positive,' indicating the system incorrectly flagged it as an error. Top right: A red sports car labeled 'sportscar' with the classification 'True Negative,' indicating the system correctly identified it as a valid generation. Bottom left: A plant labeled 'modern house' with the classification 'False Negative,' indicating the system failed to flag it as an error. Bottom right: Red flowers labeled 'person' with the classification 'True Positive,' indicating the system correctly flagged it as an error. The gray background represents cases where the system predicted a generation error.
    }
\end{figure}

Figure~\ref{fig:filter_eval} demonstrates the successes (True Positive, True Negative) and failures (False Positive, False Negative) of our filtering approach. 

\subsection{Image inpaiting pipeline parameters}
\label{sec:appendix-parameters}
We empirically tested different parameter configurations for models used in our image inpainting pipeline and set the default based on what worked best in our tests. We set \textit{strength = 1} (\ie how much the selected area should be denoised before generation), \textit{guidance = 8} (\ie how much the model follows the prompt), and \textit{step\_size = 70} (\ie the number of inference steps the model performs to generate the output image) as default parameters for the image inpainting model. We used \textit{``text, bad quality, poorly drawn, bad anatomy, blurry''} as negative prompt to discourage bad generations.

\subsection{Prompt Templates}
\subsubsection{Prompt Template for Concept Palette Clustering}
\label{sec:appendix-prompt-clustering}
This is the prompt template we used to cluster a list of tuples of alternative indices and their prompts from the entire history (\textit{e.g.}, [0] yellow flowers, [1] bushes, [2] pink flowers) into a JSON dictionary with the cluster names and the alternative indices that belong to the respective cluster (\textit{e.g.}, \{"flowers": [0, 2], ... \}):

{\scriptsize
\begin{lstlisting}
{
    "role": "system",
    "content": "Take the role of a system that is clustering imagine inpainting prompts. I will give you a list of image inpainting entries (index and prompt). You will cluster them based on content/topic similarity. You will return nothing else but a dictionary of the following format:
  
    {
        "topicName1": [0, 1, 2, 3 ...],
        "topicName2": [4, 50, 12, ...],
        ....
    }
    
    You will determine the topic name and the respective indices that belong to it. You will have an array of indices for each topic name (topic names are the keys of the dictionary.). Each index needs to be in a cluster and no index can be in multiple clusters. This means that each index will be present exactly once in your response."
},
...few-shot examples...
{
    "role": "user",
    "content": "This is my next request:
        {<@\textcolor{teal}{list\_of\_alternative\_index\_prompt\_pairs}@>}"
}
\end{lstlisting}
}

\subsubsection{Prompt Template for Concept Palette Clustering}
\label{sec:appendix-prompt-filtering}
This is the prompt template we used to filter out unsuccessful generations:

{\scriptsize
\begin{lstlisting}
{
    "role": "user",
    "content": [
        {
            "type": "text",
            "text": "Take the role of evaluating whether a prompt for image inpainting was successful. I will give you a prompt and the resulting image. You will return a one word result which is either "True" if successful or "False" if not successful. An inpainting action is determined to be successful if the subject of the prompt is visible in result image. The result image does not need to follow the prompt accurately, it is successful as long as one part of the prompt is visible in the result image (i.e. mainly the subject). This is my first request:"
        },
        {
            "type": "image_url",
            "image_url": {
                "url": "data:image/png;base64,{<@\textcolor{teal}{masked\_out\_image}@>}"
            }
        }
    ]
}
\end{lstlisting}
}


\end{document}